\documentclass[journal]{IEEEtran}
\usepackage{multirow, multicol}
\usepackage{comment, hyperref}
\usepackage{makecell, graphicx}
\usepackage{booktabs, subfigure, xcolor}
\usepackage{lineno}
\ifCLASSINFOpdf
\else
\fi

\begin{document}
%
\title{Assessment of Personality Dimensions Across Situations in Dyadic Role-Play Scenarios}
%
%
%

\author{Alice~Zhang,~\IEEEmembership{Student Member,~IEEE,}
        Skanda~Muralidhar,
        Daniel~Gatica-Perez,~\IEEEmembership{Member,~IEEE,}        
        and~Mathew~Magimai-Doss,~\IEEEmembership{Member,~IEEE}
\thanks{A. Zhang is with Idiap Research Institute, Switzerland and The University of Texas at Austin, USA. (email: alice.zhang@austin.utexas.edu)}
\thanks{D. Gatica-Perez is with Idiap Research Institute and EPFL, Switzerland.}
\thanks{S. Muralidhar and M. Magimai-Doss are with Idiap Research Institute, Switzerland.}

}

\maketitle
\begin{abstract}
Research indicates that users prefer assistive technologies whose personalities align with their own. This has sparked interest in automatic personality perception (APP), which aims to predict an individual’s perceived personality traits. Prior studies in APP have treated personalities as static traits, independent of context. However, perceived personalities can vary by situation as shown in psychological research. In this study, we investigate the relationship between conversational speech and perceived personality for participants engaged in two work situations (a neutral interview and a stressful client interaction). Our key findings are: 1) perceived personalities differ significantly across interactions, 2) loudness, sound level, and spectral flux features are indicative of perceived Extraversion, Agreeableness, Conscientiousness, and Openness in neutral interactions, while Neuroticism correlates with these features in stressful contexts, 3) handcrafted acoustic features and non-verbal features outperform speaker embeddings in inference of perceived personality, and 4) stressful interactions are more predictive of Neuroticism, aligning with existing psychological research.
\end{abstract}


%
\IEEEpeerreviewmaketitle

\vspace{-10pt}

\section{Introduction}
%
%
%
%
\IEEEPARstart{M}{odeling} human personality is fundamental to the development of affective computing systems capable of personalized interactions. Users are more engaged with and have greater trust in assistive technologies that reflect their own personalities, thereby demonstrating the value of personality-aware systems \cite{braun2019at, snyder2023busting}. Consequently, there is a growing field of research on automatic personality perception (APP), the task of inferring one's personality as perceived by external judges. Unlike automatic personality recognition (APR), which focuses on inferring self-reported personality, APP captures perceived personality and better reflects the cues that affective computing systems are designed to interpret.  


Describing an individual's personality is a complex task. The Five-Factor Model of Personality (Big-5) is a widely accepted description of personality traits in five dimensions: Extraversion, Agreeableness , Conscientiousness, Neuroticism, and Openness to Experience \cite{mccrae1992intro}. A number of previous works have leveraged this description of personality to infer traits using modalities such as speech \cite{mohammadi2015automatic, chen2017auomated}, facial expressions \cite{kachur2020assessing}, body language \cite{subramanian2013on}, and physiological signals \cite{Subramanian2018ascertain}. The APP task  is not new, as there exists a body of personality computing works \cite{Vinciarelli2014survey}; however, existing works have mainly utilized datasets in which subjects are annotated in a single context. For instance, in the widely used Speaker Personality Corpus \cite{mohammadi2015automatic, schuller2012speaker}, individuals are rated solely based on their behaviors in news broadcasts. Thus, analyses from these datasets assume that personalities are static and independent of external factors. 

On the other hand, psychology and organizational behavior researchers have long recognized that personality expression can be situation dependent \cite{mischel1968, Endler1976TowardAI, mischel1983some}.  Experts in these fields have spent decades researching and debating the stability of traits across roles and environments as seen in the personality-situation debate. While researchers have since generally acknowledged an interactionist view in which both an individual's traits and the given situation influence expression of personality through behavior, affective computing models often lack mechanisms to adjust for context \cite{Endler1976TowardAI, mischel1983some}. 

Despite the importance of studying personality in relation to situational context, computational analyses of personality across situations remain limited. Addressing this gap, we utilize the UbImpressed dataset in which participants engaged in both job-related neutral and stressful conversations and received personality ratings for both interactions \cite{muralidhar2016training}. Given the conversational nature of this dataset, we aim to pose the question: \textit{To what extent do conversational features (e.g., speech, non-verbal cues) explain perceived personality dimensions across varying contexts?} To answer this, we investigate the following research questions (RQ): 

\textbf{RQ1} Is there a significant difference in {ratings} of perceived personality of the same participants across two job-related conversation scenarios?

\textbf{RQ2} How does the relationship between perceived personality and conversational features vary across contexts?

\textbf{RQ3} How do conversational features from different interactions differ in their inference of perceived personality? 



To our knowledge, while other researchers have studied constructs such as job performance on the two situations in this dataset, ours is the first to investigate perceived personality across two different situations in this dataset \cite{muralidhar2018tale, muralidhar2018facing}. 

The paper is structured as follows: section \ref{related-work} discusses relevant work in affective computing and psychology, section \ref{dataset} describes the dataset, section \ref{methods} discusses the methodology for three analyses performed to study the research questions, section \ref{results} presents results of each analysis, and section \ref{conclusion} concludes the paper.

 

\section{Related Work}
\label{related-work}

\subsection{Contextual Variability in Expression of Personality Traits}

Researchers in psychology have long discussed the extent to which personality traits versus situational factors play a role in shaping behavior as demonstrated by the person-situation debate \cite{mischel1968}.  Proponents of the ``person" perspective argue that individuals exhibit relatively consistent behaviors across situations, suggesting that underlying traits drive these behavioral patterns. In contrast, advocates of the ``situation" perspective found that individuals' behaviors are relatively inconsistent across time and situation, arguing that personality traits do not exist and behaviors are influenced by the situation more than the individual's disposition \cite{fleeson2008end}.

While many resolutions have been proposed, a synthesis resolution acknowledging that both personality traits and situational factors influence expressed behavior is generally accepted  \cite{Endler1976TowardAI, mischel1983some, fleeson2008end}.  Over long periods of time, traits predict behavior well and explain differences in behavior between people. In momentary behavior, an individual's behavior is variable and traits may not strongly describe behaviors \cite{fleeson2004moving}. Thus, both perspectives are necessary for a full understanding of personality. However, existing works in APP have mainly focused on personality as a static trait while psychology research has shown that situational factors also influence expressed personality. For more comprehensive computing solutions to infer personality, the situation in which the solution is employed also needs to be taken into consideration.
\vspace{-10pt}

\subsection{Explaining Contextual Variability in Personality Within Job-Related Settings}
\label{explain_contextual_variability}
Two theories have been proposed to explain variations in personality-related responses across job-relevant contexts: the situation strength principle \cite{mischel1973selective} and trait activation theory (TAT) \cite{tett2003personality}. The situation strength principle proposes that strong situations, defined as environments in which rules, structures or cues provide clear guidance on an individual's expected behavior, restrict expression of personality. Meanwhile, weak situations, in which there are fewer cues regarding expected behavior, allow for greater expression of personality \cite{mischel1973selective, judge2015person}. Complementing this theory, TAT states that individuals express specific traits when situations offer opportunities for a specific trait to be expressed. For instance, an individual with high Extraversion may exhibit more extroverted behavior when working in a sales role involving customer interaction.


Complementary to research around dimensions of personality, recent psychological research has given rise to frameworks, such as DIAMONDS \cite{rauthmann2014} and CAPTIONS \cite{parrigon2017caption}, that characterize situations along a set of dimensions (e.g., \textit{complexity} - does the situation require deep thinking?).  Furthermore, dimensions of situations predict which personality traits are most readily displayed. For instance, pleasant situations most strongly allow for the manifestation of behaviors consistent with Extraversion, Agreeability, and Openness, thereby suggesting that individuals engage in more prosocial behaviors in positive situations \cite{rauthmann2014, parrigon2017caption}. 

This body of psychological work illustrates the significance of studying personality within its situational context. While prior investigations of the UbImpressed dataset used perceived personality to predict hirability and job performance \cite{muralidhar2016training, muralidhar2017how}, our work aims to predict personality from conversational behaviors and examine how this prediction task is influenced by the context of the interaction.

\vspace{-10pt}

\subsection{Automatic Personality Perception using Speech}

 As speech signals encode rich information in addition to the spoken content itself, speech is a promising modality for the APP task. For instance, the 2012 INTERSPEECH Speaker Trait Challenge \cite{schuller2012speaker} resulted in a body of different approaches towards personality inference using speech ranging from low-level acoustic descriptors to spectrum analysis \cite{pohjalainen2012feature, ivanov2012modulation}. More recent works showed that short utterance filler words \cite{tayarani2022what} and dictionary learning of spectrograms \cite{carbonneau2020feature} can be used to classify speaker's personality traits. However, prior works treat personalities as a static trait and independent of context. They utilize datasets in which only one set of personality ratings was collected per participant, whether the data came from crowd-sourced monologue interview responses \cite{chen2017auomated}, clips of video blogs from YouTube \cite{barchi2023apparent, biel2013youtube}, or speech from news clips \cite{mohammadi2015automatic, schuller2012speaker, pohjalainen2012feature, ivanov2012modulation}.  
 
In contrast to these static approaches, recent work in emotion recognition has  demonstrated the importance of contextual information \cite{kosti2017emotion, lee2019context}. However, this context-aware paradigm has not yet been fully explored in personality computing. While isolated studies have examined the relationship between personality and speech tasks (e.g., reading neutral texts versus commenting on thematic apperception test images \cite{guidi2019analysis}), there is an absence of work that compares the same individuals across varying contexts to quantify the shift in conversational behaviors for personality computing. Our study fills this gap by utilizing a paired-scenario design to isolate the effect of context on personality perception. 
\vspace{-8pt}

\section{Dataset}
\label{dataset}

We use the UbImpressed dataset \cite{muralidhar2016training}, a corpus of dyadic interactions designed to study workplace behavior. This is the only dataset to our knowledge to contain ratings for the same participants across distinct conversational contexts, allowing for intra-subject analysis of perceived personality. While newer large-scale datasets exist, they neither allow for intra-subject analysis \cite{sun-etal-2024-revealing} nor contain unscripted speech \cite{sun-etal-2024-revealing}, which are essential to our research questions. Thus, despite its age, UbImpressed is uniquely suited for this investigation. 

\vspace{-1em}

\subsection{Participants and Protocol}
The study cohort consists of 100 students (57 female, 43 male) from a hospitality management school, with a mean age of 20.6 years (SD = 2.47). Reflecting the school's bilingual curriculum, the dataset contains conversations in English (23\%) and French (77\%).
The data collection was part of a two-part behavioral training program. All 100 participants completed the first session, and 69 participants completed the second session, for a total of 169 sessions. In each session, students engaged in two role-play scenarios:
\begin{enumerate}
\item \textbf{Employment interview:} The student acted as an applicant for a hospitality internship, while a research assistant played the recruiter. Questions focused on motivation and experience. While job interviews can be stressful, this one was framed as a practice opportunity. 
\item \textbf{Hotel reception desk interaction with an unsatisfied customer:} The student acted as a hotel receptionist handling a client (played by a research assistant), who was unsatisfied with charges on their bill and hostile. This interaction was confrontational and more stressful compared to the \textit{interview}. The increased stress of this interaction is empirically verified in our analysis of personality ratings (section \ref{rating_analysis_results}). 
\end{enumerate}

Following the first session, students received feedback from professionals in human resources (HR) on their performance. The dataset contains a total of 338 interactions (169 sessions x 2 scenarios). Although the scenarios were simulated and the interlocutor's speech was scripted, the participants' speech was entirely spontaneous. The average duration of the \textit{interview} was 7.6 minutes and 4.6 minutes for the \textit{desk}. From here on, we refer to \textit{scenario} as the \textit{interview} or \textit{desk} interaction and the \textit{session} as the first (1) or second (2) instance of an interaction. 
\vspace{-2em}
\subsection{Perceived Personality Rating}
Perceived personality was manually annotated based on the Big-5 traits and stress.  Annotators were Master's students in the same psychology program, ensuring a shared understanding of the trait definitions. The rating process was performed using 2-minute slices of the audiovisual recording, a common practice in psychology \cite{Ambady1992ThinSlices}. To minimize bias between scenarios, the \textit{interview} and \textit{desk} interactions were annotated by independent groups of annotators (five raters for \textit{interviews}; three for \textit{desk} interactions). No annotator rated the same participant in both scenarios. Each participant received a score on a Likert-scale of 1 (Low) to 7 (High) for each dimension. As shown in Table \ref{tab:icc}, inter-rater reliability (assessed via Intraclass Correlation Coefficient, ICC(2, k)) indicated at least moderate Agreement (ICC $\geq$ 0.5) for most traits. 
\vspace{-7pt}

\begin{table}
    \scriptsize
    \centering
    \caption{Intra-class correlation for perceived personalities and stress.}
     \begin{tabular}{c|c|c|c|c|c|c|c} \hline
     \multirow{2}*{Session} & \multirow{2}*{Scenario} & \multicolumn{6}{c}{\textit{ICC(2,k)}}\\ \cline{3-8}
     & & Extra & Agree & Consc & Neuro & Open & Stress \\ \hline
     \multirow{2}*{1} & Interview & 0.65& 0.54& 0.59& 0.51& 0.48 & 0.74\\
      & Desk & 0.62& 0.72& 0.69& 0.56& 0.33 & 0.63\\    \hline
     \multirow{2}*{2} & Interview & 0.66& 0.50& 0.56& 0.45& 0.37 & 0.54\\
      & Desk & 0.77& 0.58& 0.59& 0.37& 0.58 & 0.45\\
     \end{tabular}
     \label{tab:icc}
\vspace{-10pt}
\end{table}

\section{Methods and Experimental Setup}
\label{methods}

\begin{figure*}[!ht]
    \centering
    \includegraphics[width=0.9\linewidth]{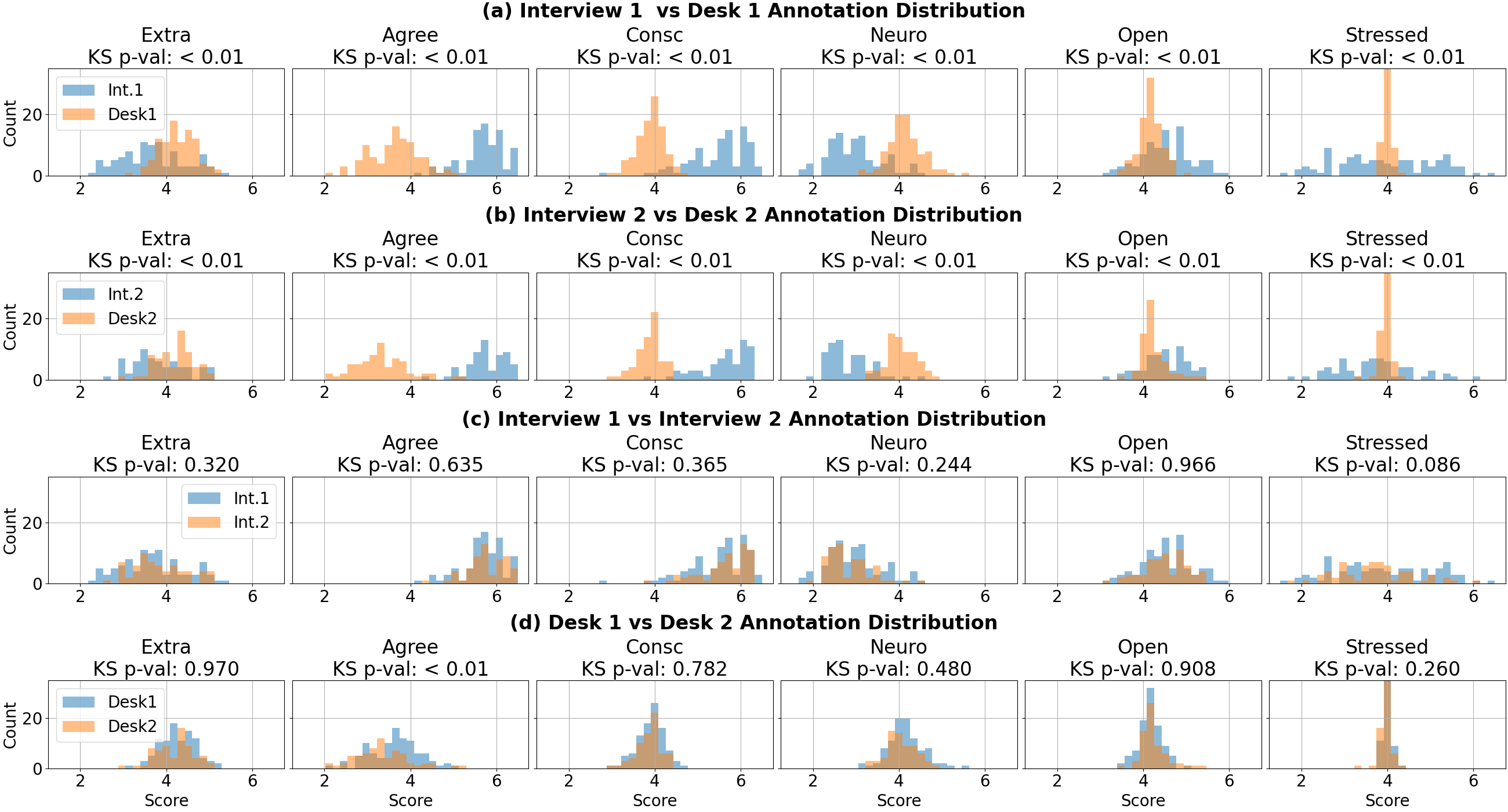}
    \caption{(RQ1) Comparing distribution of perceived personality scores between: (a) interview and desk scenario of first lab session, (b) interview and desk scenario of second lab session, (c) interview scenario of first and second lab sessions, and (d) desk scenario of first and second lab sessions.}
    \label{fig:personality-diff}
    \vspace{-10pt}
\end{figure*}
\subsection{Personality Rating Comparison across Conversations}
\label{rating-comparison}
To understand whether perceived personality differs across conversations (RQ1), we analyze the distribution of ratings across the \textit{interview} and \textit{desk} scenarios and the first and second sessions of each interaction. We employ the two-sample Kolmogorov-Smirnov (KS) test for goodness of fit where the null hypothesis states that the underlying continuous distributions $F(x)$ and $G(x)$ of two independent samples are identical for all $x$ \cite{Hodges1958TheSP}. We first use the two-sample KS test to compare the distributions of participants' perceived stress across different scenarios and sessions to verify that the two interaction scenarios in the UbImpressed dataset are significantly different to validate the use of the UbImpressed dataset for our research questions. Then, we use the two-sample KS test to compare the distribution of perceived personality across different scenarios and sessions to answer RQ1.

\vspace{-8pt}


\subsection{Feature Extraction and Selection}
We diarize each conversation and keep only the audio segments corresponding to speech from the students. From each student's speech, we extract features that have been shown to vary with a speaker's emotional state and influenced by personality traits \cite{mairesse2007using}.  While recent advancements in speech processing heavily utilize deep learning embeddings to maximize accuracy, these representations lack interpretability. Because our objective is to identify how specific behavioral mechanisms drive personality perception under varying conversational contexts, we prioritize interpretable feature sets. Consequently, we select (1) eGeMAPS features and (2) non-verbal features. To contextualize the performance of these handcrafted features against newer representations, we also benchmark them against (3) speaker embeddings. The specific dimensions corresponding to each category are available \href{https://osf.io/8vztd/?view_only=57e6b2ae1602457a89bf9472889a4d64}{here}\footnote{\url{https://osf.io/8vztd/?view_only=57e6b2ae1602457a89bf9472889a4d64}\label{fn:link}}.

\textbf{eGeMAPS} \cite{eyben2015geneva} features are a set of acoustic features commonly used for speech-based affective computing using the openSMILE toolkit \cite{eyben2010opensmile}. It contains 88 features that capture frequency, energy, amplitude, and spectral parameters initially hand-crafted for speech emotion recognition. 

\textbf{Non-verbal features} include 75 audio and visual cues selected for their relevance in existing literature in psychology \cite{DeGrootGooty2009} and social computing \cite{nguyen2014hire}, and previously extracted for the UbImpressed dataset. We utilize this set because it provides a multimodal behavioral profile of a speaker that eGeMAPS features alone cannot address. The feature categories include: speaking activity (21 dimensions; e.g. turn duration, pause statistics), prosody (30 dimensions; e.g., spectral entropy, voicing statistics), head nods (8 dimensions), visual back-channeling (6 dimensions) and overall visual motion (10 dimensions). These features fundamentally capture conversation-level dynamics and are thus extracted at the conversation-level.

\textbf{Speaker embeddings} are fixed-dimensional representations of speech that encode speaker identity. Work by Ulgen et \textit{al.} \cite{ulgen2024revealing} show that intra-speaker clusters of ECAPA-TDNN vectors \cite{desplanques2020ecapa}, a type of speaker embedding, can capture affective states. Thus, we use 512-dimensional ECAPA-TDNN vectors extracted via the Pyannote toolkit \cite{Bredin2020pyannote}. 


We extract eGeMAPS and speaker embedding features at the utterance (turn) level, following prior literature \cite{eyben2015geneva, desplanques2020ecapa}. To model the overall interaction, we then aggregated these turn-level features into a single, conversation-level vector by calculating the mean or median of each feature dimension across a participant's conversation. We found that personality inference using the median feature vector consistently outperformed the mean aggregation. Thus, our primary analyses utilize this median vector to represent the interaction.

To investigate the temporal dynamics of personality expression and capture how acoustic markers evolve over the course of an interaction, we conducted an additional temporal analysis with the eGeMAPS features, since they generally outperformed speaker embeddings. We isolated feature vectors from five distinct temporal landmarks based on the chronological sequence of each participant's turns: the first utterance, the $25^{th}$  percentile utterance, the midpoint ($50^{th}$ percentile utterance), the $75^{th}$ percentile utterance, and the last utterance.

On the other hand, the non-verbal feature vector, which contains audio and video-based features, captures conversation-level features of the participant's interaction. Therefore, there is only one feature vector extracted for each participant per interaction, and we use the feature vector as is.

Lastly, for each feature set, we select a subset of features using the Pearson Correlation Coefficient (PCC). The PCC measures the strength of the linear relationship between two continuous variables. We assume that relevant information is contained in features significantly correlated with perceived personality scores and select features with \textit{p} $<$ 0.05 (RQ2). 

\label{feature-selection}

\vspace{-8pt}

\subsection{Experimental Setup for Inference of Perceived Personality }

We formulate the inference of perceived personality from speech signals as a regression task (RQ3) to compare personality inference across conversations. 

\subsubsection{Conversation-Specific Inference} First, we aim to infer perceived personality on a specific conversation scenario and session to understand how feature selection impacts performance. We perform feature selection as detailed in section \ref{feature-selection} using data from the complementary session of the same scenario so that no data from the test set influences the feature selection process. For instance, to build and evaluate a system inferring personality in the first \textit{interview} session, we perform feature selection using data from the second \textit{interview} session.

To infer a continuous score for each personality dimension, we train regression models using one of the three feature sets. Our primary architecture is a Random Forest (RF) Regressor with 200 trees, depth of 5, and 16 samples per leaf node. We selected a RF because it captures non-linear relationships and provides interpretable feature importance metrics. To ensure the robustness of our model selection, we benchmarked the RF against ElasticNet, K-Nearest Neighbors (KNN) and Support Vector Machine regressors and conducted hyperparameter optimization (e.g., tuning maximum depth for the RF) of all models to ensure fair comparison. The RF consistently provided competitive or superior predictive performance across the Big-5 traits, empirically justifying its use as the primary model for our downstream feature importance analysis. 

We evaluate all regressors using the coefficient of determination $R^2$, which is the amount of total variance explained by the model, and the PCC between the predicted and observed personality scores in a 10-fold cross validation scheme. Each feature vector corresponds to one participant, so each evaluation fold consists of speakers unseen in the training process. Thus, the evaluation is speaker independent. We obtain confidence intervals by bootstrapping the evaluation set.
\label{conv-specific}
\vspace{-10pt}
\subsubsection{Inference across Scenarios and Sessions} To understand the generalization of personality regression across different scenarios within the same set of participants, we build a RF regressor trained on data from one scenario (e.g. \textit{interview 1}) and evaluate using data from the complementary scenario of the same session (e.g. \textit{desk 1}). We perform feature selection using the same method as in Sections \ref{feature-selection} and \ref{conv-specific}. 

Then, to understand how a trained RF regressor generalizes across conversation instances of the same scenario over time, we build a regressor trained on data from one session (e.g. \textit{interview 1}) and evaluate it using data from the complementary session of the same scenario (e.g. \textit{interview 2}). We perform feature selection using the training dataset. In this setup, the training and testing sets are not strictly participant-independent, as the 69 participants who returned for the second session appear in both sets. However, because each session was annotated separately and participants modified their behavior based on HR feedback, both behavior and target labels differ across sessions. This setup evaluates the model's capacity to generalize to behavioral shifts within a cohort (see Figure 1 in Supplementary Material).

\vspace{-6pt}

\section{Results and Analysis}
\label{results}

\subsection{Personality Rating Comparison Results and Analysis}
\vspace{-3pt}

\label{rating_analysis_results}
As shown in Figure \ref{fig:personality-diff}, we reject the null hypothesis of the two-sample KS test in favor of the alternative hypothesis when comparing all personality dimensions and stress scores across \textit{interview} and \textit{desk} scenarios. In contrast, we fail to reject the null hypothesis in comparing personality and stress across the first and second sessions of the \textit{interview} or \textit{desk} scenario, with the exception of Agreement in the \textit{desk} scenario. This analysis shows that participants are perceived to exhibit significantly more stress in the client interaction ($\mu =$ 4.06) compared to the interview ($\mu =$  3.82) and validates the suitability of the UbImpressed dataset to study our research questions.

Overall, this analysis shows that perceived personality differs significantly between the neutral interview and stressful client interaction for the same speaker (RQ1). In the \textit{interview}, participants were perceived as more agreeable ($\mu =$  5.81 vs. 3.53) and open ($\mu =$  4.52 vs. 4.19), and less neurotic ($\mu =$  2.92 vs. 4.11). While psychological frameworks have established that personality expression varies by situation \cite{rauthmann2014, parrigon2017caption}, these studies rely on participants' retrospective self-reports of past events. Our results provide empirical validation that observers perceive these personality shifts from interactions.  

\vspace{-9pt}

\subsection{Correlation Analysis Results and Analysis}
\vspace{-3pt}

\label{corr-analysis}
\begin{table}[!h]
    \scriptsize
    \centering
    \caption{(RQ2) Correlation of select eGeMAPS and non-verbal features with personality. \textsuperscript{\dag} $p$ $<$ 0.01; \textsuperscript{*} $p$ $<$ 0.05.}.
     \begin{tabular}{c|c|c|c|c|c|c} \hline
     Feature & Scenario & Extra & Agree & Consc & Neuro & Open \\ \hline
    
     \multirow{4}*{\makecell{\textit{equivalent}\\\textit{Sound}\\\textit{Level (dB)}\\(eGeMAPS)}} & Int. 1 & 0.52\textsuperscript{\dag} & 0.43\textsuperscript{\dag} & 0.31\textsuperscript{\dag} & -0.08 & 0.41\textsuperscript{\dag} \\ 
     & Int. 2 & 0.41\textsuperscript{\dag} & 0.34\textsuperscript{*} & 0.37\textsuperscript{\dag} & -0.27\textsuperscript{*} & 0.27\textsuperscript{*} \\
     & Desk 1 & 0.09 & 0.41\textsuperscript{\dag} & 0.04 & -0.36\textsuperscript{\dag} & 0.19 \\
     & Desk 2 & 0.21 & 0.49\textsuperscript{\dag} & 0.15 & -0.27\textsuperscript{*} & 0.07 \\ \hline

     
     \multirow{4}*{\makecell{\textit{\# pauses}\\\textit{\textgreater 1 second}\\(non-verbal)}} & Int. 1 & 0.18 & 0.23\textsuperscript{*} & 0.16 & -0.07 & -0.27\textsuperscript{*} \\ 
     & Int. 2 & 0.31\textsuperscript{*} & 0.22 & 0.04 & -0.13 & 0.38\textsuperscript{\dag} \\
     & Desk 1 & -0.12 & 0.23\textsuperscript{*} & -0.03 & -0.27\textsuperscript{*} & 0.24\textsuperscript{*} \\
     & Desk 2 & 0.07 & 0.15 & 0.07 & -0.02 & 0.02 \\ \hline

     \multirow{4}*{\makecell{\textit{Count}\\\textit{Head Nods}\\(non-verbal)}} & Int. 1 & 0.42\textsuperscript{\dag} & 0.25\textsuperscript{*} & 0.24\textsuperscript{*} & -0.06 & 0.31\textsuperscript{\dag} \\ 
     & Int. 2 & 0.44\textsuperscript{\dag} & 0.01 & -0.07 & -0.07 & 0.29\textsuperscript{*} \\
     & Desk 1 & 0.25\textsuperscript{*} & 0.07 & 0.10 & -0.24\textsuperscript{*} & 0.08 \\
     & Desk 2 & 0.39\textsuperscript{\dag} & 0.22 & 0.04 & -0.30\textsuperscript{*} & 0.09 \\ 
     
     \end{tabular}
     \label{tab:egemaps-corr}
    \vspace{-5pt}
     
\end{table}

Select results of the correlation between each participant's median eGeMAPS and non-verbal feature vector, and personality dimensions are presented in Table \ref{tab:egemaps-corr} (RQ2). The complete correlation matrix for all feature sets can be found \href{https://osf.io/8vztd/?view_only=57e6b2ae1602457a89bf9472889a4d64}{here} \footref{fn:link}.

For both sessions, there are stronger correlations between eGeMAPS and nonverbal features and personality in the \textit{interview} for all personality dimensions except Neuroticism, in which the correlation between Neuroticism and features is stronger in the \textit{desk} scenario. Additionally, there is an inverse in correlation direction between Neuroticism and features. We hypothesize this is due to the association between Neuroticism and negative emotions such as anxiety, in which prior research has shown that anxiety introduces irregularities in conversational behaviors compared to positive or neutral speech \cite{ozseven2018voice}. 


For eGeMAPS features, there is a common subset of features that are significantly correlated with all personality dimensions except Neuroticism in the \textit{interview}. Energy- (loudness and equivalent sound level) and spectral- (spectral flux) related features are all correlated with perceived Extraversion, Agreement, Conscientiousness, and Openness in the \textit{interview}.  In contrast, the correlation is significant between this subset of features and Neuroticism only in the \textit{desk} scenario. Interestingly, significant correlations between this subset of speech features and Agreement exist in both the \textit{interview} and \textit{desk} interaction, suggesting that this trait may be more consistently expressed across contexts. 

For non-verbal features, we observe that features related to speech energy, voicing rate, and head nods are positively correlated with all personality dimensions except Neuroticism in the \textit{interview}. In the \textit{desk} scenario, we observe positive correlations between features related to speaking activity, number of pauses, and speech energy, and Agreement but negative correlations between these features and Neuroticism. 
\vspace{-2em}

\begin{table*}
\caption{(RQ3) Regression results for predicting personality dimensions within a given scenario and session. Cells with a dimension (Dims) of features set to 0 indicate there are no features with statistically significant correlations with the personality trait. Bold highlights the best performance for the given personality dimension across both scenarios. \textsuperscript{{\dag}}$p$ $<$ 0.01; \textsuperscript{*}$p$ $<$ 0.05. }

    \tiny
    \centering
    
     \begin{tabular}{c|c|c|c|c|c|c|c|c|c|c|c} \hline
     \multirow{2}*{Feature} & \multirow{2}*{Session} & \multicolumn{2}{c|}{Extraversion} & \multicolumn{2}{c|}{Agreement} & \multicolumn{2}{c|}{Conscientiousness} & \multicolumn{2}{c|}{Neuroticism} & \multicolumn{2}{c}{Openness} \\
     &  & Dims & $R^2$, $r$ & Dims & $R^2$, $r$ & Dims & $R^2$, $r$ & Dims & $R^2$, $r$ & Dims & $R^2$, $r$ \\ \hline
     
     \multirow{4}*{Hand-crafted (eGeMAPS)} & \multirow{2}*{1} &  88 & 0.25 $\pm$ 0.14\textsuperscript{*}, 0.51\textsuperscript{\dag} & 88 & 0.16 $\pm$ 0.13\textsuperscript{*}, 0.41\textsuperscript{\dag}& 88 & 0.07 $\pm$ 0.10, 0.28 & 88 & -0.04 $\pm$ 0.10, 0.07 & 88 & 0.16 $\pm$ 0.12, 0.39\textsuperscript{\dag}\\
     &  &  27 & 0.27 $\pm$ 0.14\textsuperscript{*}, 0.52\textsuperscript{\dag} & 52 & 0.19 $\pm$ 0.14\textsuperscript{*}, 0.42\textsuperscript{\dag}& 39 & 0.01 $\pm$ 0.13, 0.19 & 15 & -0.05 $\pm$ 0.12, 0.05 & 18 & 0.19 $\pm$ 0.15\textsuperscript{*}, 0.42\textsuperscript{\dag}\\ \cline{2-12}
     
    & \multirow{2}*{2}  & 88 & 0.02 $\pm$ 0.17, 0.22 & 88& 0.18 $\pm$ 0.12\textsuperscript{*}, 0.45\textsuperscript{\dag}& \textbf{88} & \textbf{0.24 $\pm$ 0.13, 0.51\textsuperscript{*}}& 88 & -0.02 $\pm$ 0.10, 0.11 & 88 & -0.08 $\pm$ 0.15, 0.03\\
     &  & 31 & 0.03 $\pm$ 0.17, 0.25\textsuperscript{*} & \textbf{43} & \textbf{0.20 $\pm$ 0.12\textsuperscript{*}, 0.47\textsuperscript{\dag}}& 28 & 0.18 $\pm$ 0.13, 0.47\textsuperscript{\dag} & 6 & -0.08 $\pm$ 0.11, 0.01 & 32 & -0.06 $\pm$ 0.16, 0.09\\ \hline

    \multirow{4}*{Non-verbal} & \multirow{2}*{1} &  75  & 0.29 $\pm$ 0.12, 0.55\textsuperscript{\dag} & 75 & 0.18 $\pm$ 0.13 \textsuperscript{*}, 0.41\textsuperscript{\dag} & 75 & 0.07 $\pm$ 0.11, 0.27\textsuperscript{\dag} & 75 & -0.05 $\pm$ 0.09, 0.02 & \textbf{75} & \textbf{0.22 $\pm$ 0.12, 0.46\textsuperscript{\dag}} \\
     
     & & \textbf{35} & \textbf{0.30 $\pm$ 0.12\textsuperscript{*}, 0.55\textsuperscript{\dag}} & 23 & 0.19 $\pm$ 0.13\textsuperscript{*}, 0.43\textsuperscript{\dag} & 15 & 0.06 $\pm$ 0.12, 0.26\textsuperscript{\dag} & 2 & -0.05 $\pm$ 0.09, -0.01 & 18 & 0.20 $\pm$ 0.14, 0.44\textsuperscript{\dag} \\ \cline{2-12}
     
     & \multirow{2}*{2} & 75 & 0.17 $\pm$ 0.11\textsuperscript{*}, 0.55\textsuperscript{\dag} & 75 & 0.07 $\pm$ 0.11, 0.44\textsuperscript{\dag} & 75 & 0.10 $\pm$ 0.15, 0.34\textsuperscript{\dag} & 75 & -0.03 $\pm$ 0.09, 0.08 & 75 & 0.04 $\pm$ 0.11, 0.24\textsuperscript{*}\\
      &  & 32 & 0.12 $\pm$ 0.15\textsuperscript{*}, 0.38\textsuperscript{\dag} & 35 & 0.08 $\pm$ 0.11, 0.32\textsuperscript{\dag} & 28 & 0.10 $\pm$ 0.17, 0.35\textsuperscript{\dag} & 6 & -0.07 $\pm$ 0.08, -0.08 & 36 & 0.06 $\pm$ 0.12, 0.28\textsuperscript{*} \\ \hline

     

    \end{tabular}
    
    \vspace{0.8em}
    \centerline{\small (a) Interview}

    \vspace{1.5em}


     \begin{tabular}{c|c|c|c|c|c|c|c|c|c|c|c} \hline
     \multirow{2}*{Feature Type} & \multirow{2}*{Session} & \multicolumn{2}{c|}{Extraversion} & \multicolumn{2}{c|}{Agreement} & \multicolumn{2}{c|}{Conscientiousness} & \multicolumn{2}{c|}{Neuroticism} & \multicolumn{2}{c}{Openness} \\
     &  & Dims & $R^2$, $r$ & Dims & $R^2$, $r$ & Dims & $R^2$, $r$ & Dims & $R^2$, $r$ & Dims & $R^2$, $r$ \\ \hline   
     \multirow{4}*{Hand-crafted (eGeMAPS)} & \multirow{2}*{1} & 88 & -0.04 $\pm$ 0.08, 0.02 & 88 & 0.08 $\pm$ 0.11\textsuperscript{*}, 0.30\textsuperscript{*} & 88 & -0.10 $\pm$ 0.07, -0.16 & 88 & 0.08 $\pm$ 0.10, 0.31\textsuperscript{\dag} & 88 & -0.10 $\pm$ 0.10, -0.09\\
     
     &  &  20 & -0.07 $\pm$ 0.07, -0.11 & 26 & 0.09 $\pm$ 0.12, 0.33\textsuperscript{\dag} & 0 & - & \textbf{13} & \textbf{0.12 $\pm$ 0.14, 0.36\textsuperscript{\dag}} & 1 & -0.07 $\pm$ 0.07, -0.03\\ \cline{2-12}
     
     & \multirow{2}*{2} & 88 & -0.03 $\pm$ 0.17, 0.13 & 88 & 0.18 $\pm$ 0.16, 0.44\textsuperscript{\dag} & 88 & -0.15 $\pm$ 0.08, -0.39\textsuperscript{\dag} & 88 & -0.07 $\pm$ 0.09, -0.01 & 88 & -0.14 $\pm$ 0.09, -0.23\\
     &  &  5 & -0.09 $\pm$ 0.05, -0.38\textsuperscript{\dag} & 24 & 0.18 $\pm$ 0.16, 0.45\textsuperscript{\dag} & 2 & -0.09 $\pm$ 0.05, -0.34\textsuperscript{\dag} & 21 & -0.01 $\pm$ 0.12, 0.16 & 11 & -0.15 $\pm$ 0.08, -0.35\textsuperscript{\dag}\\ \hline

    \multirow{4}*{Non-verbal} & \multirow{2}*{1} &  75 & -0.01 $\pm$ 0.08, 0.09  & 75 & 0.01 $\pm$ 0.10, 0.20 & 75 & -0.03 $\pm$ 0.08, 0.06 & 75 & 0.07 $\pm$ 0.10, 0.28\textsuperscript{\dag} & 75 & -0.11 $\pm$ 0.08, -0.10\\
    
     &  & 10 & 0.02 $\pm$ 0.10, 0.20 & 16 & 0.01 $\pm$ 0.10, 0.23\textsuperscript{*} & 1 & -0.09 $\pm$ 0.04, -0.27\textsuperscript{\dag} & 13 & 0.02 $\pm$ 0.10, 0.20 &  2 & -0.05 $\pm$ 0.08, 0.01 \\ \cline{2-12}
     
     & \multirow{2}*{2} & 75 & -0.04 $\pm$ 0.08, 0.01  & 75 & -0.07 $\pm$ 0.10, -0.02 & 75 & -0.10 $\pm$ 0.08, -0.18 & 75 & 0.02 $\pm$ 0.13, 0.22 & 75 & -0.12 $\pm$ 0.08, -0.21\\
     
      &  & 13 & 0.02 $\pm$ 0.10, 0.18 & 30 & -0.02 $\pm$ 0.11, 0.10 &  3 & -0.11 $\pm$ 0.05, -0.51\textsuperscript{\dag} & 28 & 0.05 $\pm$ 0.12, 0.28\textsuperscript{*} & 7 & -0.14 $\pm$ 0.09, -0.43\textsuperscript{\dag} \\ \hline

     
     \end{tabular}
     
    \vspace{0.8em}
    \centerline{\small (b) Desk}
    \label{tab:personality-prediction}
    \vspace{-10pt}
    
\end{table*}

\begin{figure*}
    \centering
    \includegraphics[width=0.85\textwidth]{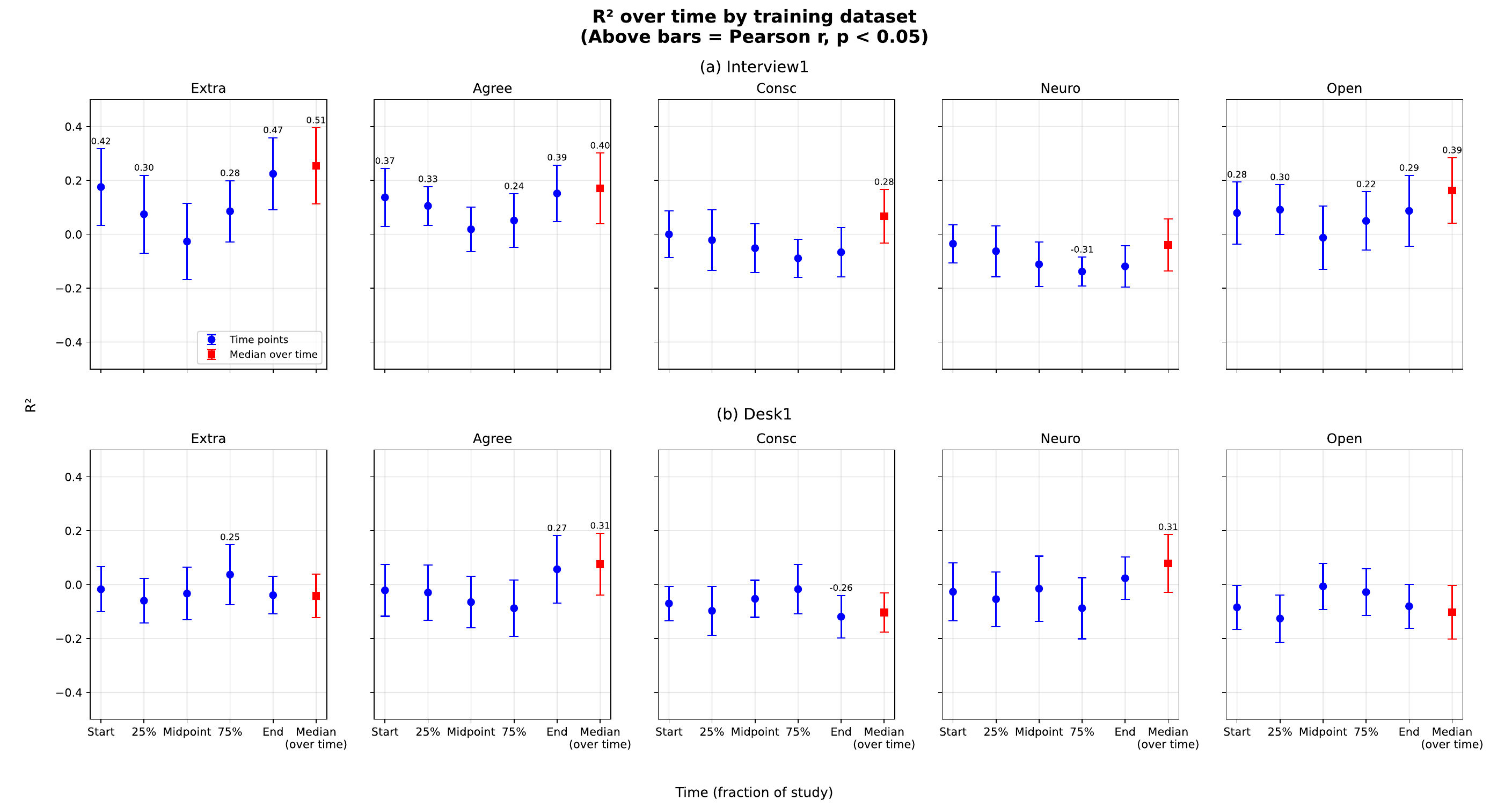}
    \caption{Temporal dynamics of predicting personality in (a) Interview 1 and (b) Desk 1.}
    \label{fig:results_over_time}
    \vspace{-10pt}

\end{figure*}

\subsection{Inference of Perceived Personality Results}

\subsubsection{Conversation-Specific Perceived Personality}
\label{conversation-specific-prediction}

We observe that the RF regressor contains some predictive power for all personality dimensions across all three feature sets (Table \ref{tab:personality-prediction}; eGeMAPs and non-verbal results only, speaker embedding results in Supplementary Material). Generally, eGeMAPS and non-verbal features outperform speaker embeddings, with non-verbal features explaining a maximum 30\% of the variance for the Extraversion dimension. Our eGeMAPS results are comparable to the results obtained by Barchi \textit{et al.} who report $R^2$ values between 0.12 and 0.22 for each personality dimension \cite{barchi2023apparent}. We prioritize this comparison because \cite{barchi2023apparent} isolates the performance of eGeMAPS features on spontaneous speech. To our knowledge, no comparable works on personality regression using non-verbal and speaker embedding features exist. 

Our temporal analysis reveals different dynamics in the  contexts for session 1 (Figure \ref{fig:results_over_time}, see Supplementary Material for session 2). In the \textit{interview}, the predictive performance is relatively stable across time. The median feature vector generally achieves the most robust performance across traits (notably for Extraversion, Agreeableness, and Conscientiousness in session 1). We also observe that the first and last utterances yield localized peaks in performance for Extraversion and Agreeableness in the first session. This aligns with first and last impressions often setting the tone of an interaction and thus carrying concentrated social signals. In contrast, for the \textit{desk} scenario, the temporal dynamics were more volatile. For traits like Agreement, Neuroticism, and Openness, some utterances from temporal landmarks consistently outperformed the median aggregate. These results show that personality expression in stressful scenarios is inconsistent over time and markers of personality appear at specific moments.  

Our regressor performs comparably with and without feature selection, showing that a subset of features can have as much predictive power as the entire feature set. Additionally, across both sessions and all feature sets, we observe that the \textit{desk} dataset is better for inference of Neuroticism and that the \textit{interview} dataset is better for inference of the other four personality dimensions. These results support the correlation results presented in section \ref{corr-analysis} and also psychological research on the inference of personality from situations. Situations in which negative feelings can occur and lack enjoyment are associated with Neuroticism whereas situations containing pleasant interactions are associated with Extraversion, Agreement, and Openness \cite{rauthmann2014, parrigon2017caption}. Additionally, stressful situations are more likely to  activate the expression of Neuroticism \cite{tett2003personality}.

\begin{table*}[!htbp]
    \tiny
    \centering
    \caption{(RQ3) Regression results for eGeMAPS. Predicting personality dimensions when training on one scenario and evaluating on the other scenario within the same session. Cells with a dimension (Dims) of features set to 0 indicate that there are no features with statistically significant correlations to the personality dimension. \textsuperscript{{\dag}}$p$ $<$ 0.01; \textsuperscript{*}$p$ $<$ 0.05.}
     \begin{tabular}{c|c|c|c|c|c|c|c|c|c|c|c|c} \hline
     \multirow{2}*{Session} & \multirow{2}*{\makecell{Train\\Scenario}} & \multirow{2}*{\makecell{Eval.\\Scenario}} & \multicolumn{2}{c|}{Extraversion} & \multicolumn{2}{c|}{Agreement} & \multicolumn{2}{c|}{Conscientiousness} & \multicolumn{2}{c|}{Neuroticism} & \multicolumn{2}{c}{Openness} \\
     &  &  & Dims & $R^2$, $r$ & Dims & $R^2$, $r$ & Dims & $R^2$, $r$ & Dims & $R^2$, $r$ & Dims & $R^2$, $r$ \\ \hline
     
     \multirow{4}*{1} & \multirow{2}*{Interview} & \multirow{2}*{Desk}  & 88 & -0.41 $\pm$ 0.29, 0.17 & 88 & -19.1 $\pm$ 5.0, 0.11 & 88 & -32.0 $\pm$ 8.1, 0.07 & 88 & -8.6 $\pm$ 2.8, 0.12 & 88 & -7.7 $\pm$ 2.0, 0.14\\
     &  &  & 27 & -0.64 $\pm$ 0.34, 0.11 & 52 & -19.9 $\pm$ 5.3, 0.13 & 39 & -32.3 $\pm$ 7.9, -0.01 & 15 & -8.2 $\pm$ 2.4, -0.11 & 18 & -8.5 $\pm$ 2.3, 0.22\textsuperscript{*}\\ \cline{2-13}
     
     & \multirow{2}*{Desk} & \multirow{2}*{Interview} & 88 & -0.47 $\pm$ 0.26, 0.27\textsuperscript{\dag} & 88 & -20.5 $\pm$ 5.6, 0.31\textsuperscript{\dag} & 88 & -6.0 $\pm$ 2.1, 0.07 & 88 & -4.8 $\pm$ 1.7, 0.01 & 88 & -0.4 $\pm$ 0.22, 0.25\textsuperscript{*}\\
     & &  &  20& -0.40 $\pm$ 0.28, 0.40\textsuperscript{\dag}& 26 & -20.5 $\pm$ 5.6, 0.22\textsuperscript{*} & 0 & - & 13 & -4.7 $\pm$ 1.6, 0.13 & 1 & -0.33 $\pm$ 0.22, 0.17\\ \hline 
     
     \multirow{4}*{2} & \multirow{2}*{Interview} & \multirow{2}*{Desk} & 88 & -0.55 $\pm$ 0.43, 0.3\textsuperscript{*} & 88 & -20.9 $\pm$ 8.8, 0.21 & 88 & -41.6 $\pm$ 11.5, -0.01 & 88 & -16.1 $\pm$ 4.7, -0.07 & 88 & -2.6 $\pm$ 1.3, -0.11\\
     &  &  & 31 & -0.75 $\pm$ 0.50, 0.29\textsuperscript{*} & 43 & -20.9 $\pm$ 8.9, 0.22 & 28 & -36.4 $\pm$ 10.0, 0.13& 6 & -17.0 $\pm$ 5.0, -0.04 & 32 & -3.1 $\pm$ 1.4, 0.02 \\ \cline{2-13}     
     
     & \multirow{2}*{Desk} & \multirow{2}*{Interview} & 88 & -0.09 $\pm$ 0.15, 0.29\textsuperscript{*} & 88 & -35.5 $\pm$ 10.9, 0.04 & 88 & -11.8 $\pm$ 4.5, -0.21 & 88 & -7.33 $\pm$ 3.4, -0.03 & 88 & -0.40 $\pm$ 0.28, 0.01\\
     & &  & 5 & -0.29 $\pm$ 0.22, 0.10 & 24& -35.2 $\pm$ 10.8, 0.33\textsuperscript{\dag}& 2 & -10.9 $\pm$ 4.1, -0.02& 21 & -7.3 $\pm$ 3.4, -0.08 & 11 & -0.43 $\pm$ 0.29, 0.16\\ \hline
     \end{tabular}

    \label{tab:same-session}
    \vspace{-8pt}

\end{table*}
\begin{table*}[!htbp]
    \tiny
    \centering
    \caption{(RQ3) Regression results for eGeMAPS. Predicting personality dimensions when training on one session and evaluating on the other session of the same scenario.  \textsuperscript{{\dag}}$p$ $<$ 0.01; \textsuperscript{*}$p$ $<$ 0.05.} 
     \begin{tabular}{c|c|c|c|c|c|c|c|c|c|c|c|c} \hline
     \multirow{2}*{Scenario} & \multirow{2}*{\makecell{Train\\Session}} & \multirow{2}*{\makecell{Eval.\\Session}} & \multicolumn{2}{c|}{Extraversion} & \multicolumn{2}{c|}{Agreement} & \multicolumn{2}{c|}{Conscientiousness} & \multicolumn{2}{c|}{Neuroticism} & \multicolumn{2}{c}{Openness} \\
     &  &  & Dims & $R^2$, $r$ & Dims & $R^2$, $r$ & Dims & $R^2$, $r$ & Dims & $R^2$, $r$ & Dims & $R^2$, $r$ \\ \hline
     
     \multirow{4}*{Interview} & \multirow{2}*{1} & \multirow{2}*{2} & 88 & 0.09 $\pm$ 0.28\textsuperscript{\dag}, 0.41\textsuperscript{\dag}& 88 & 0.26 $\pm$ 0.16\textsuperscript{\dag}, 0.51\textsuperscript{\dag}& 88 & 0.18 $\pm$ 0.15\textsuperscript{\dag}, 0.53\textsuperscript{\dag}& 88 & -0.032 $\pm$ 0.12\textsuperscript{\dag}, 0.21 & 88 & 0.05 $\pm$ 0.17\textsuperscript{\dag}, 0.30\textsuperscript{*}\\
     
     &  &  & 31 & 0.10 $\pm$ 0.28\textsuperscript{\dag}, 0.41\textsuperscript{\dag}& 43& 0.27 $\pm$ 0.16\textsuperscript{\dag}, 0.52\textsuperscript{\dag}& 28 & 0.10 $\pm$ 0.18\textsuperscript{\dag}, 0.43\textsuperscript{\dag}& 6 & -0.14 $\pm$ 0.18, 0.10 & 32 & 0.03 $\pm$ 0.18\textsuperscript{\dag}, 0.30\textsuperscript{*}\\ \cline{2-13}
     
     & \multirow{2}*{2} & \multirow{2}*{1} & 88& 0.17 $\pm$ 0.09\textsuperscript{\dag}, 0.56\textsuperscript{\dag}& 88 & 0.17 $\pm$ 0.10\textsuperscript{\dag}, 0.49\textsuperscript{\dag}& 88 & 0.07 $\pm$ 0.12\textsuperscript{\dag}, 0.34\textsuperscript{\dag}& 88 & -0.02 $\pm$ 0.13, 0.20\textsuperscript{*} & 88 & 0.11 $\pm$ 0.06\textsuperscript{\dag}, 0.41\textsuperscript{\dag}\\
     
     &  &  & 27 & 0.19 $\pm$ 0.08\textsuperscript{\dag}, 0.57\textsuperscript{\dag}& 52 & 0.17 $\pm$ 0.10\textsuperscript{\dag}, 0.48\textsuperscript{\dag}& 39 & 0.07 $\pm$ 0.13, 0.34\textsuperscript{\dag}& 15 & -0.02 $\pm$ 0.14, 0.19 & 18 & 0.15 $\pm$ 0.08\textsuperscript{\dag}, 0.45\textsuperscript{\dag}\\ \hline

     \multirow{4}*{Desk} & \multirow{2}*{1} & \multirow{2}*{2} & 88 & -0.04 $\pm$ 0.13, 0.14 & 88 & -0.21 $\pm$ 0.36, 0.46\textsuperscript{\dag}& 88 & -0.08 $\pm$ 0.11, 0.02 & 88 & 0.01 $\pm$ 0.25\textsuperscript{\dag}, 0.29\textsuperscript{*}& 88 & -0.05 $\pm$ 0.12, 0.09\\
     
     &  &  & 5 & -0.09 $\pm$ 0.20, 0.10 & 24 & -0.21 $\pm$ 0.37, 0.48\textsuperscript{\dag}& 2 & -0.14 $\pm$ 0.18, -0.03 & 21 & -0.07 $\pm$ 0.26\textsuperscript{*}, 0.29\textsuperscript{*}& 11 & -0.09 $\pm$ 0.11, 0.03\\ \cline{2-13}

     & \multirow{2}*{2} & \multirow{2}*{1} & 88 & -0.11 $\pm$ 0.21, 0.08 & 88 & -0.33 $\pm$ 0.34, 0.40 \textsuperscript{\dag}& 88 & -0.06 $\pm$ 0.09, 0.05 & 88 & 0.03 $\pm$ 0.08\textsuperscript{\dag}, 0.42\textsuperscript{\dag} & 88 & -0.04 $\pm$ 0.07, 0.03\\
     
     &  &  & 20 & -0.18 $\pm$ 0.18, 0.03 & 26 & -0.32 $\pm$ 0.34, 0.42\textsuperscript{\dag}& 0 & - & 13 & 0.11 $\pm$ 0.08\textsuperscript{\dag}, 0.50\textsuperscript{\dag}& 1 & -0.05 $\pm$ 0.10, 0.12\\ \hline        
     \end{tabular}

    \label{tab:same-scenario}
    \vspace{-8pt}
     
\end{table*}    

\subsubsection{Perceived Personality Across Scenarios and Sessions}

We report results of personality inference on the scenario unseen during training in Table \ref{tab:same-session} (eGeMAPS only; others in Supplementary Material). The regressor performs significantly worse in this cross-scenario validation despite the training and evaluation datasets containing the same participants. For most personality dimensions, the features do not explain any of the variance. Only non-verbal features explain a minimal amount of variance in Extraversion when training on the \textit{interview} and evaluating on the \textit{desk} interaction, indicating that certain non-verbal behaviors are predictive of Extraversion in both stressful and neutral interactions. But, the overall lack of generalizability emphasizes that the relationship between speech and personality varies across situations.

The results of personality inference across sessions of the same scenario are in Table \ref{tab:same-scenario} (eGeMAPS only; others in Supplementary Material). For eGeMAPS features, this evaluation setup outperforms the evaluation in which the training and test data come from the same session and scenario. The performance improvement is perhaps due to training with the entire conversation dataset rather than nine of ten folds. 

We still observe that the \textit{desk} interaction is more predictive of Neuroticism. These results combined with the results from the cross-scenario evaluation highlight that, within a type of scenario, there are specific and relatively consistent relationships between speech features and perceived personality dimensions. However, the relationships do not generalize well across different conversation scenarios. In conclusion, our results demonstrate that acoustic and non-verbal markers of personality are highly context-dependent; thus, affective computing systems should move beyond static models to account for situational context and temporal dynamics of the interaction.

\section{Limitations and Future Work}

\subsection{Ecological Validity of Role-Play}
\vspace{-3pt}
Our findings are specific to goal-oriented, dyadic interactions within a professional setting due to our use of simulated role-play interactions. However, while the interaction context was fixed, the participants' speech was unscripted and spontaneous. Furthermore, HR professionals provided performance feedback, ensuring the interactions carried meaning for the participants. Future work should investigate whether these conversational patterns persist in unrestricted, naturalistic settings.

Additionally, our analyses rely on the UbImpressed dataset and represent a specific demographic (hospitality management students). Consequently, the generalizability of our results to other populations remains to be verified. Nonetheless, our results highlight the importance of modeling context to better perceive and adapt to users. For instance, in an application for clinical interviews, distinguishing between situational anxiety and Neuroticism as a trait would be critical for diagnosis.

\vspace{-5pt}
\subsection{Real-World Applicability}
While our modest predictive power ($R^2$) limits immediate real-world application, explainable variance must be evaluated against the theoretical upper bound of inter-rater agreement \cite{nili2014toolbox}. For example, given an Extraversion ICC of 0.65, our model ($R^2 = 0.30$) effectively captures 46\% ($0.30 / 0.65$) of the explainable variance. Ultimately, our objective is not achieving state-of-the-art accuracy, but using interpretable features to analyze personality inference shifts across contexts.

\vspace{-10pt}

\subsection{Speech Emotion Recognition} This study focuses on neutral and stressful contexts to isolate the effect of situational stress on personality perception. Future studies could integrate speech emotion recognition (SER) to distinguish between specific emotional states within each context. Such an analysis would disentangle whether emotional cues drive differences in conversational behavior observed in personality perception and prediction \cite{greenaway2018context}.
\vspace{-8pt}

\section{Conclusion}
\label{conclusion}
We investigated the relationship between perceived personality and conversational speech using the UbImpressed dataset. Our experiments showed that perceived personality of the same participants differs significantly across a neutral and stressful interaction \textbf{(RQ1)}. We found that features related to loudness, equivalent sound level, and spectral flux are correlated with perceived Extraversion, Agreement, Conscientiousness, and Openness in the neutral scenario (\textit{interview}) and with perceived Neuroticism in the stressful scenario (\textit{desk interaction}) \textbf{(RQ2)}. Through a regression analysis, we demonstrated that eGeMAPS and non-verbal features explain up to 27\% and 30\% of variance respectively in perceived Extraversion. Furthermore, features better predicted Neuroticism in the \textit{desk} scenario, but the remaining personality dimensions in the \textit{interview} scenario \textbf{(RQ3)}. Our results emphasize the relevance of building and evaluating APP systems relative to the context in which behaviors are observed. Furthermore, our results highlight the importance of context-aware affective computing systems, which has thus far been understudied.
\vspace{-8pt}

\section*{Acknowledgment}
The authors gratefully acknowledge the Swiss Federal Commission for Scholarships for funding this research with the Swiss Government Excellence Scholarship, grant 2024.0241 and Innosuisse Flagship project IICT,
grant PFFS-21-47. 

\vspace{-8pt}
\ifCLASSOPTIONcaptionsoff
  \newpage
\fi



%
\bibliography{mybib}


%
\bibliographystyle{IEEEtran}
\vspace{-4em}

\begin{IEEEbiographynophoto}{Alice Zhang} is a Ph.D. candidate at the University of Texas at Austin. Her research interests include acoustic sensing and wearable technologies to support and analyze social interactions.
\end{IEEEbiographynophoto}
\vspace{-5em}

\begin{IEEEbiographynophoto}{Skanda Muralidhar} (Ph.D. EPFL) is a researcher at Idiap Research Institute, where he researches and develops AI-powered technologies for sign language learning. His current work advances applied AI for education, healthcare, and human-centered innovation.
\end{IEEEbiographynophoto}
\vspace{-5em}

\begin{IEEEbiographynophoto}{Daniel Gatica-Perez} directs the Social Computing Group at Idiap Research Institute and is a professor at EPFL. His research integrates human-centered and participatory methods with mobile, social, and AI technologies to support individuals and communities.
\end{IEEEbiographynophoto}
\vspace{-5em}

\begin{IEEEbiographynophoto}{Mathew Magimai-Doss} (Ph.D. EPFL) is a Senior Researcher at Idiap Research Institute, Switzerland. His research interests lie in signal processing, statistical pattern recognition, artificial neural networks and computational linguistics with applications to speech, audio, and multimodal signal processing. \end{IEEEbiographynophoto}
\vspace{-4em}



\end{document}